\documentclass[5p,twocolumn,authoryear]{elsarticle}

\usepackage{natbib}
\biboptions{authoryear,round}
\usepackage{amsmath,amssymb}
\usepackage{graphicx}

\usepackage{hyperref}
\usepackage{booktabs}
\usepackage{float}
\usepackage{placeins}
\usepackage{caption}
\usepackage{subcaption}

\journal{NeuroImage}

\begin{document}

\begin{frontmatter}

\title{A Dynamical Microscope for Multivariate Oscillatory Signals: Validating Regime Recovery on Shared Manifolds}

\author[1]{Łukasz Furman\corref{cor1}}
\author[2,3]{Ludovico Minati}
\author[1]{Włodzisław Duch}

\address[1]{Faculty of Physics, Astronomy and Informatics, Nicolaus Copernicus University, 87-100, Toruń, Poland}
\address[2]{School of Life Science and Technology, University of Electronic Science and Technology of China, 611731 Chengdu, China}
\address[3]{Center for Mind/Brain Sciences (CIMeC), University of Trento, 38123 Trento, Italy}

\cortext[cor1]{Corresponding author}

\begin{abstract}
Multivariate oscillatory signals from complex systems often exhibit non-stationary dynamics and metastable regime structure, making dynamical interpretation challenging.
We introduce a ``dynamical microscope'' framework that converts multichannel signals into circular phase--amplitude features, learns a data-driven latent trajectory representation with an autoencoder, and quantifies dynamical regimes through trajectory geometry and flow field metrics.
Using a coupled Stuart--Landau oscillator network with topology-switching as ground-truth validation, we demonstrate that the framework recovers differences in dynamical laws even when regimes occupy overlapping regions of state space.
Group differences can be expressed as changes in latent trajectory speed, path geometry, and flow organization on a shared manifold, rather than requiring discrete state separation.
Speed and explored variance show strong regime discriminability ($\eta^2 > 0.5$), while some metrics (e.g., tortuosity) capture trajectory geometry orthogonal to topology contrasts.
The framework provides a principled approach for analyzing regime structure in multivariate time series from neural, physiological, or physical systems.
\end{abstract}


\end{frontmatter}

\section{Introduction}

\subsection{Motivation: multivariate signals as dynamical systems}

Complex systems---whether neural, physiological, or physical---often produce multivariate oscillatory signals that are highly dynamic and metastable. Unlike responses time-locked to external events, ongoing activity exhibits continual fluctuations and short-lived patterns of coordination among components. Subsystems spontaneously couple and decouple in time, reflecting a balance between integration and segregation \citep{TognoliKelso2014}. Metastability implies that the system does not reside in a single stationary state but transitions among a repertoire of semi-stable configurations across multiple time scales \citep{Rabinovich2008,TognoliKelso2014}. These observations reinforce that ongoing activity is inherently non-stationary and structured in time.

Characterizing such dynamics is challenging because continuous recordings lack external markers that naturally segment the signal. Segment choice is often arbitrary, and stationarity assumptions are frequently violated. Traditional stationary summaries (e.g., a single spectrum or connectivity matrix over a long epoch) can obscure transient structure. These issues motivate approaches that treat multivariate signals as continuous trajectories evolving in a high-dimensional state space, where the objects of interest are not isolated states but the geometry and flow of trajectories over time.

Importantly, for such signals the ``trajectory'' should not be understood as a single clean curve in latent space. A more appropriate picture is a \emph{stochastic flow} on a low-dimensional manifold: at each moment, the latent state undergoes structured drift (local flow) together with substantial variability, yielding an evolving probability density over states. As a result, single short realizations can appear as ``snake-like'' paths, while pooling longer time spans or multiple realizations naturally produces dense point clouds with metastable basins (high-occupancy regions) and preferred directions of motion (flow fields). In this sense, trajectory-centric (Lagrangian) and density/field-centric (Eulerian) views are complementary descriptions of the same underlying dynamics.

\subsection{From ``state labels'' to trajectory laws}

Many frameworks for analyzing dynamics reduce continuous activity to a sequence of discrete ``states.'' In microstate analysis, each time point is assigned to one of a few prototypical patterns, and comparisons across conditions focus on metrics such as fractional occupancy and mean duration \citep{Michel2018}. Similarly, hidden Markov model approaches represent activity as probabilistic switching among latent states with characteristic patterns and dwell times \citep{Baker2014,Vidaurre2018}. Such state-based methods are useful for summarizing recurring configurations, but a primary focus on state occupancy can miss differences that are expressed in the \emph{motion} through state space.

A complementary hypothesis is that conditions or regimes may share a common manifold (i.e., occupy broadly similar regions of state space) yet differ systematically in their trajectory laws: local flow structure, velocities, path geometry, recurrence, and exploration versus confinement. Theoretical work links altered dynamical conditions to changes in metastability and temporal exploration rather than wholesale changes in the set of accessible configurations \citep{Cavanna2018,Deco2017}. This motivates shifting emphasis from ``which states are visited'' to ``how trajectories evolve,'' and from static labels to dynamical descriptions of motion on learned coordinates.

\subsection{Continuity with existing methods and new contributions}

Our framework is designed to complement, not replace, discrete-state methods. Some quantities we compute---such as occupancy entropy and dwell statistics---have direct analogues in microstate and HMM analyses (e.g., state entropy, transition rates, mean durations). We include these for continuity and interpretability, not as claims of novelty.

The novel contribution lies in \emph{continuous-trajectory quantities} that do not reduce cleanly to discrete-state summaries:
\begin{itemize}
    \item \textbf{Path tortuosity}: a geometric functional of the continuous trajectory (curvature and zig-zag structure) that captures within-region motion, not just boundary crossings between states.
    \item \textbf{Speed variability (CV)}: local kinematics variability that quantifies how intermittently fast or slow the system moves, beyond simple ``more or fewer switches.''
    \item \textbf{Flow field geometry}: spatially resolved drift structure showing \emph{where} and \emph{how} trajectories move, not just which regions are occupied.
\end{itemize}
These quantities are not directly recoverable from standard discrete-state summaries without additional assumptions, and require explicit continuous-trajectory representations.

\subsection{Summary of contributions}

We adopt a trajectory-centric dynamical-systems view and introduce a framework designed as a ``dynamical microscope'' rather than a classifier. Our contributions are:
\begin{itemize}
    \item A circular phase--amplitude encoding of multichannel signals that respects phase geometry and captures amplitude modulations.
    \item A latent-trajectory framework in which an autoencoder learns a compact representation of ongoing dynamics, treated explicitly as continuous trajectories.
    \item Trajectory- and flow-based metrics that quantify dynamical regimes in terms of motion, geometry, and exploration---complementing discrete-state summaries (entropy, dwell) with continuous-trajectory quantities (tortuosity, speed variability, flow geometry) not directly recoverable from standard discrete-state summaries.
    \item Ground-truth validation on a coupled Stuart--Landau oscillator network with topology-switching, demonstrating sensitivity to differences in dynamical laws even when state-space regions overlap.
    \item Systematic analysis of metric sensitivity across different regime contrasts, with speed and variance showing strong discriminability ($\eta^2 > 0.5$).
\end{itemize}

\paragraph{A note on the role of the autoencoder.}
The autoencoder serves as a nonlinear coordinate transformation, not a model of the underlying generative process; dynamical quantities are computed \emph{post hoc} and are not optimized during training (see Methods). We emphasize that the autoencoder is not essential to the framework; any smooth coordinate chart preserving temporal continuity (e.g., PCA, diffusion maps) could be substituted. The autoencoder is used here as a flexible nonlinear chart that improves local flow coherence, not as a generative model.

\begin{figure*}[t]
    \centering
    \includegraphics[width=0.95\textwidth]{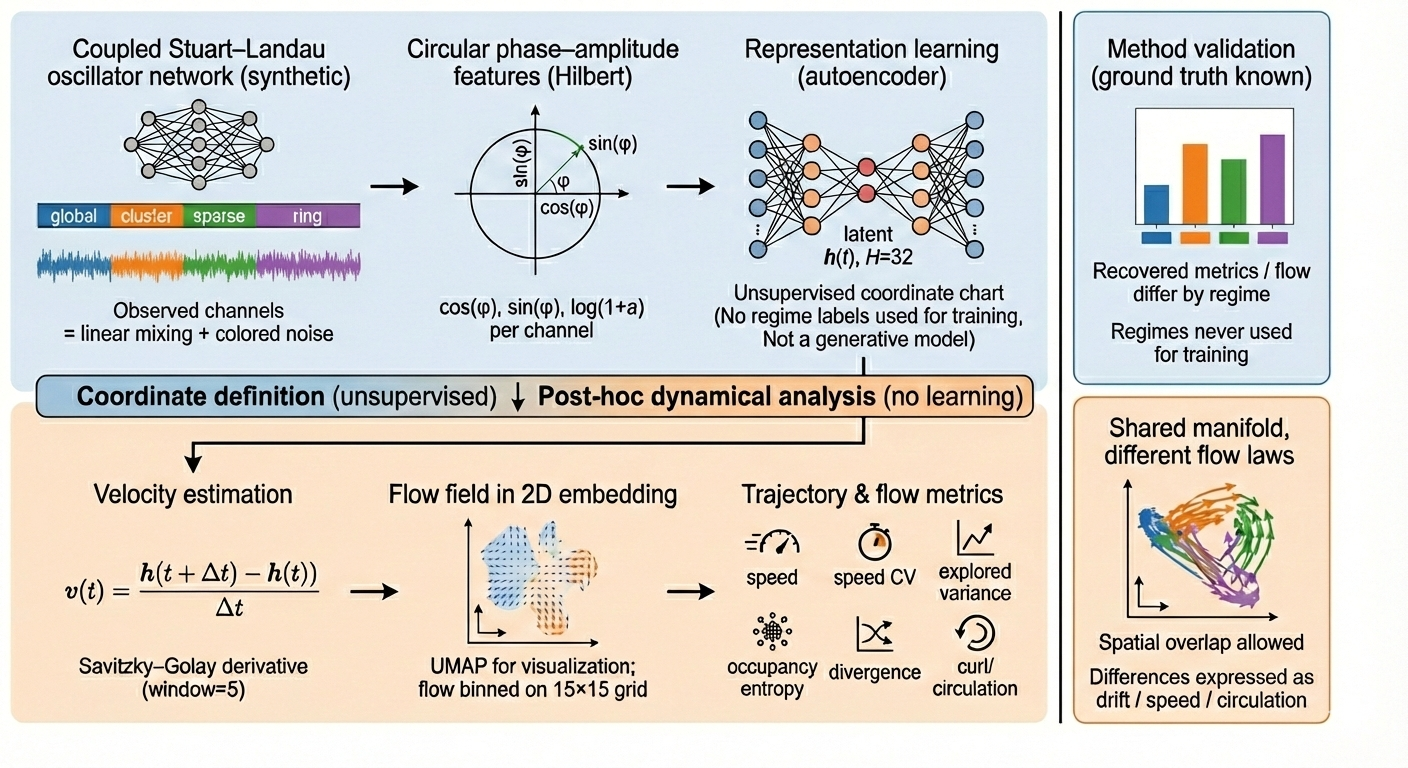}
    \caption{Workflow schematic of the dynamical microscope framework. Multichannel oscillatory signals are transformed into circular phase--amplitude features, encoded into a latent trajectory via autoencoder, and analyzed using flow field estimation and trajectory metrics. The depicted ``trajectory'' is schematic; analyses emphasize both time-ordered realizations and ensemble summaries (densities and flow fields) obtained by pooling across time and/or realizations.}
    \label{fig:workflow}
\end{figure*}

\section{Related Work}

\subsection{Dynamical systems views of complex signals}

Dynamical systems theory provides a principled framework for understanding complex signals as trajectories evolving on structured state spaces \citep{Kelso1995,Breakspear2017}. A central concept is metastability: transient coordination among subsystems that preserves both integration and flexibility \citep{TognoliKelso2014,Rabinovich2008}. Computational models suggest that rich switching among configurations emerges near critical regimes, yielding broad dynamical repertoires \citep{Deco2017}. These perspectives motivate trajectory-centric descriptions that quantify how activity moves through state space.

\subsection{Representation learning for time series}

Deep learning has enabled representation learning for multivariate time series, including self-supervised approaches seeking transferable representations \citep{Banville2021,Kostas2021}. However, most methods are evaluated by predictive utility, with latent spaces treated as features rather than dynamical state spaces. In contrast, we use representation learning as a coordinate transformation preserving moment-to-moment variation, producing continuous trajectories for dynamical analysis rather than classification \citep{Hinton2006,CunninghamYu2014}.

Related approaches combine latent-variable representation learning with explicit temporal or dynamical structure, including sequential autoencoders and switching dynamical systems \citep{Pandarinath2018,Linderman2017rSLDS}, as well as generative latent-variable models that can be extended with temporal priors \citep{KingmaWelling2014}.

\subsection{Trajectory-based and recurrence-based analyses}

Nonlinear time-series methods such as recurrence quantification analysis (RQA) summarize revisitation structure in reconstructed state spaces \citep{Marwan2007}. State-space geometry approaches include microstates \citep{Michel2018} and HMM-based analyses \citep{Baker2014,Vidaurre2018}, which provide compact summaries but impose categorical boundaries. Related approaches recover dynamical laws directly from data \citep{Brunton2016,Schmid2010DMD,Chen2018}.

Our framework is complementary: it retains continuous trajectories and estimates flow fields directly in learned coordinates, enabling quantification of motion differences (speed, tortuosity, exploration) even when conditions share a manifold \citep{TognoliKelso2014,Deco2017}.

\section{Methods}

\subsection{Coupled Stuart--Landau oscillator network}
\label{sec:synthetic_data}

To validate whether the proposed dynamical microscope recovers known dynamical structure, we constructed a synthetic multivariate dynamical system with explicitly defined latent dynamics and regime structure. The simulation provides full access to ground-truth regime identity and dynamical parameters, while producing observations through a generic linear mixing model.

\subsubsection{Stuart--Landau oscillator dynamics}

The Stuart--Landau equation describes the normal form of a Hopf bifurcation and provides a canonical model for limit-cycle oscillators \citep{Strogatz1994}:
\begin{equation}
\dot{z}_j = (\mu_j + i\omega_j)z_j - |z_j|^2 z_j + \sigma\xi_j(t) + \kappa \sum_{k} L_{jk} z_k,
\end{equation}
where $z_j \in \mathbb{C}$ is the complex state of oscillator $j$, $\mu_j$ controls the limit-cycle amplitude, $\omega_j$ is the natural frequency, $\sigma\xi_j(t)$ represents additive complex noise, and $L_{jk}$ is the graph Laplacian encoding coupling topology with strength $\kappa$.

This formulation produces multivariate oscillatory dynamics where:
\begin{itemize}
    \item Each oscillator exhibits sustained oscillations on a limit cycle
    \item Coupling topology determines collective synchronization patterns
    \item Different topologies produce qualitatively different dynamical regimes
\end{itemize}

\subsubsection{Topology-switching regime structure}

We implement four distinct coupling topologies that produce contrasting dynamical organizations:

\paragraph{Global (all-to-all).} Complete graph coupling promotes global synchronization, where all oscillators tend toward phase alignment.

\paragraph{Cluster (modular).} Block-structured adjacency with dense within-cluster and sparse between-cluster connections ($p_{\text{in}}=1.0$, $p_{\text{out}}=0.01$) produces multi-cluster synchrony patterns.

\paragraph{Sparse (random).} Very sparse random connectivity (density $\approx 0.03$) promotes desynchronization and independent oscillator dynamics.

\paragraph{Ring (directional).} Directed ring lattice with nearest-neighbor coupling promotes traveling wave patterns through spatially arranged frequency gradients.

Regimes switch every 10 seconds, cycling through all four topologies four times over a 160-second recording (16 total regime segments). Transitions are smoothed using sigmoid interpolation over 300~ms to avoid discontinuities.

\subsubsection{Observation model}

Latent oscillator states are mapped to $C=30$ observed channels via a fixed random mixing matrix $\mathbf{M} \in \mathbb{R}^{C \times N_{\text{osc}}}$:
\begin{equation}
\mathbf{x}(t) = \mathbf{M}\,\text{Re}(\mathbf{z}(t)) + \boldsymbol{\eta}(t),
\end{equation}
where $\boldsymbol{\eta}(t)$ represents 1/f-colored observation noise (pink noise, $\alpha=1.0$) for EEG-like spectral characteristics. Ground-truth regime identity is recorded for evaluation and visualization only, and is not used in representation learning or metric definitions.

\subsubsection{Simulation parameters}

Table~\ref{tab:simulation_params} summarizes the simulation configuration.

\begin{table}[h]
\centering
\caption{\textbf{Coupled Stuart--Landau simulation parameters.}}
\label{tab:simulation_params}
\small
\begin{tabular}{ll}
\toprule
\textbf{Parameter} & \textbf{Value} \\
\midrule
Number of oscillators & 30 \\
Number of channels & 30 \\
Sampling rate & 250~Hz \\
Total duration & 160~s \\
Regime duration & 10~s \\
Number of cycles & 4 \\
Coupling strength $\kappa$ & 5.0 \\
Oscillator noise $\sigma$ & 0.1 \\
Observation noise & 0.05 (pink, $\alpha=1.0$) \\
Transition smoothing & 0.3~s \\
\bottomrule
\end{tabular}
\end{table}

\begin{figure}[h]
    \centering
    \includegraphics[width=0.48\textwidth]{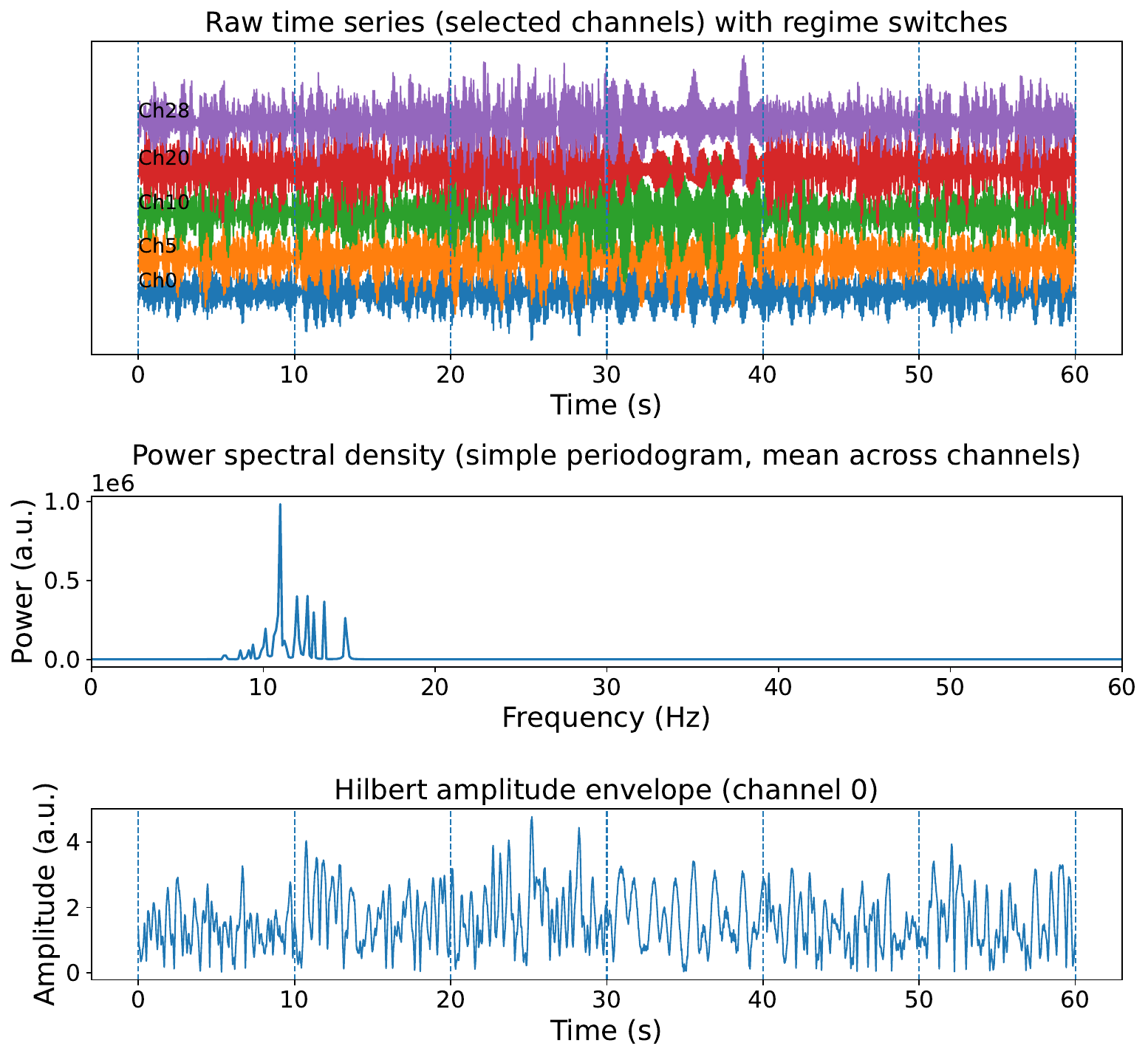}
    \caption{\textbf{Raw observations from coupled oscillator simulation.} Top: multichannel time series (selected channels) showing regime switches as vertical dashed lines. Middle: power spectral density (mean across channels) with characteristic oscillator peak. Bottom: Hilbert amplitude envelope (channel 0) showing amplitude modulations across regimes.}
    \label{fig:electrode_demo}
\end{figure}

Figure~\ref{fig:electrode_demo} illustrates the raw observations from the coupled oscillator network, showing how regime switches manifest in the multichannel time series and spectral characteristics.

\subsection{Preprocessing and circular phase--amplitude representation}

Let $\mathbf{x}(t) \in \mathbb{R}^{C}$ denote the multichannel signal. After bandpass filtering (1--30~Hz, 4th-order Butterworth, zero-phase via \texttt{filtfilt}), the analytic signal for each channel is obtained via the Hilbert transform applied to the continuous signal, yielding instantaneous phase $\phi_c(t)$ and amplitude $a_c(t)$ \citep{Boashash1992}.

Instantaneous phase and amplitude are widely used to quantify oscillatory coordination and cross-frequency interactions, and provide a natural starting point for trajectory-centric representations \citep{Canolty2006,Tort2010}.

Each channel is represented with three features:
\begin{equation}
\mathbf{z}_c(t) = \big[\cos\phi_c(t),\; \sin\phi_c(t),\; \log(1 + a_c(t))\big],
\end{equation}
and stacking channels yields the full feature vector $\mathbf{z}(t) \in \mathbb{R}^{3C}$. This representation encodes phase on the unit circle and amplitude on a logarithmic scale, avoiding phase wrap discontinuities while capturing amplitude modulations without dominating the geometry. Unless otherwise stated, all analyses use the combined phase--amplitude representation.

\subsection{Autoencoder for latent trajectories}

An autoencoder maps $\mathbf{z}(t)$ to a latent state $\mathbf{h}(t) \in \mathbb{R}^{H}$ at each time point. The model is trained with a reconstruction objective:
\begin{equation}
\mathcal{L}_{\mathrm{recon}} = \frac{1}{T}\sum_{t=1}^{T}\|\hat{\mathbf{z}}(t) - \mathbf{z}(t)\|_2^2,
\end{equation}
with no regime labels. The encoder output is interpreted as a continuous latent trajectory over time rather than a single static code.

\paragraph{Autoencoder as a coordinate chart.}
Training uses pooled data from all realizations with a reconstruction-only objective (no regime labels), yielding a locally smooth representation that preserves moment-to-moment continuity. Dynamical quantities (velocity, dwell structure, flow fields) are computed \emph{post hoc} from the learned coordinates and are not optimized during training. The autoencoder provides a nonlinear coordinate chart that improves flow-field coherence and spatial interpretability relative to linear projections.

\subsection{Embedding, flow field estimation, and dynamical metrics}

\paragraph{Terminology.}
To avoid ambiguity, we define the following terms used throughout:
\begin{itemize}
    \item \textbf{Latent code} $\mathbf{h}(t) \in \mathbb{R}^{H}$: the autoencoder's bottleneck output at each time point (dimension $H=32$).
    \item \textbf{Visualization embedding} $\mathbf{y}(t) \in \mathbb{R}^{d}$: a low-dimensional projection of $\mathbf{h}(t)$ for plotting and pooled analysis (typically $d=2$), obtained via UMAP \citep{McInnes2018}.
    \item \textbf{Flow field}: the spatially binned average of velocity vectors $\mathbf{v}(t)$ in the visualization embedding, representing local drift direction and magnitude.
\end{itemize}
Primary scalar metrics (mean speed, tortuosity, explored variance) are computed in the \emph{full latent space} $\mathbf{h}(t)$ to ensure coordinate invariance, while flow field visualizations use the 2D projection $\mathbf{y}(t)$ for interpretability.

\paragraph{Velocity estimation.}
Velocity in a sampled trajectory is inherently scale-dependent: fine-scale estimates are dominated by high-frequency noise, while coarser scales capture mesoscale flow. We use Savitzky--Golay derivative estimation \citep{SavitzkyGolay1964} with window 5 and order 2; effect direction is consistent across alternative estimators (Appendix~\ref{app:velocity_robustness}).

\subsubsection{Flow field estimation}
\label{sec:flow_field_estimation}
Flow fields are estimated by spatially binning $\mathbf{y}(t)$ and averaging velocity vectors within each bin, yielding a coarse-grained approximation of local dynamical flow. Concretely:
\begin{itemize}
    \item The 2D embedding space is discretized into a $15 \times 15$ regular grid.
    \item Bins with fewer than $k=3$ trajectory points are excluded to ensure reliable local estimates.
    \item Within each bin, the mean velocity vector $\bar{\mathbf{v}}_{\mathrm{bin}}$ is computed by averaging all $\mathbf{v}(t)$ whose corresponding $\mathbf{y}(t)$ falls in that bin.
    \item Divergence and curl are computed via finite differences on the resulting vector field:
    \begin{align}
    \nabla \cdot \mathbf{v} &\approx \frac{\partial v_x}{\partial x} + \frac{\partial v_y}{\partial y}, \\
    (\nabla \times \mathbf{v})_z &\approx \frac{\partial v_y}{\partial x} - \frac{\partial v_x}{\partial y}.
    \end{align}
\end{itemize}

From trajectories and flow fields, we compute complementary metrics that characterize motion rather than discrete state assignments:
\begin{itemize}
    \item \textbf{Mean speed} and \textbf{speed variability (CV)}.
    \item \textbf{Path tortuosity} (curvature proxy) as a measure of directional instability.
    \item \textbf{Explored variance} as a measure of state-space coverage.
    \item \textbf{Occupancy entropy} quantifying the uniformity of spatial distribution.
    \item \textbf{Divergence} and \textbf{curl} characterizing local flow field structure.
\end{itemize}

\paragraph{Intrinsic metrics.}
Primary metrics (speed, tortuosity, explored variance) are computed in the full latent space $\mathbf{h}(t) \in \mathbb{R}^{32}$, not in low-dimensional projections. This ensures that observed differences reflect intrinsic dynamical structure rather than projection artifacts.

\subsection{Statistical analysis}

Regime discriminability is assessed using one-way ANOVA across the four topology regimes, with effect size quantified by eta-squared ($\eta^2$):
\begin{equation}
\eta^2 = \frac{SS_{\text{between}}}{SS_{\text{total}}},
\end{equation}
where $\eta^2 > 0.14$ is commonly treated as a large effect, $0.06$--$0.14$ as medium, and $< 0.06$ as small \citep{Cohen1988}.

Metrics are computed on non-overlapping windows (50 samples $\approx$ 0.2~s after 4$\times$ compression) to reduce short-timescale autocorrelation and provide approximately independent windows for descriptive discriminability analysis. This windowing strategy is used to reduce short-timescale autocorrelation rather than to claim strict statistical independence. Effect sizes ($\eta^2$) are reported as primary summaries rather than $p$-values alone.

\section{Results: Validation on Coupled Oscillator Network}

\subsection{Simulation design and evaluation approach}

The coupled Stuart--Landau oscillator network was simulated with topology-switching every 10 seconds, cycling through global, cluster, sparse, and ring topologies four times (160~s total). The resulting 30-channel observations were processed through the complete pipeline (filtering, Hilbert phase--amplitude extraction, autoencoder embedding, UMAP visualization, flow field estimation, and metric computation).

Ground-truth regime labels were used only for visualization and for stratifying summary statistics; they were not used in model training, embedding, flow estimation, or metric definitions. The central question is whether the dynamical microscope can recover known regime differences when those differences are expressed in the \emph{coupling topology} rather than in positional separation.

\subsection{Regime recovery from latent trajectories}

\begin{figure}[h]
    \centering
    \includegraphics[width=0.48\textwidth]{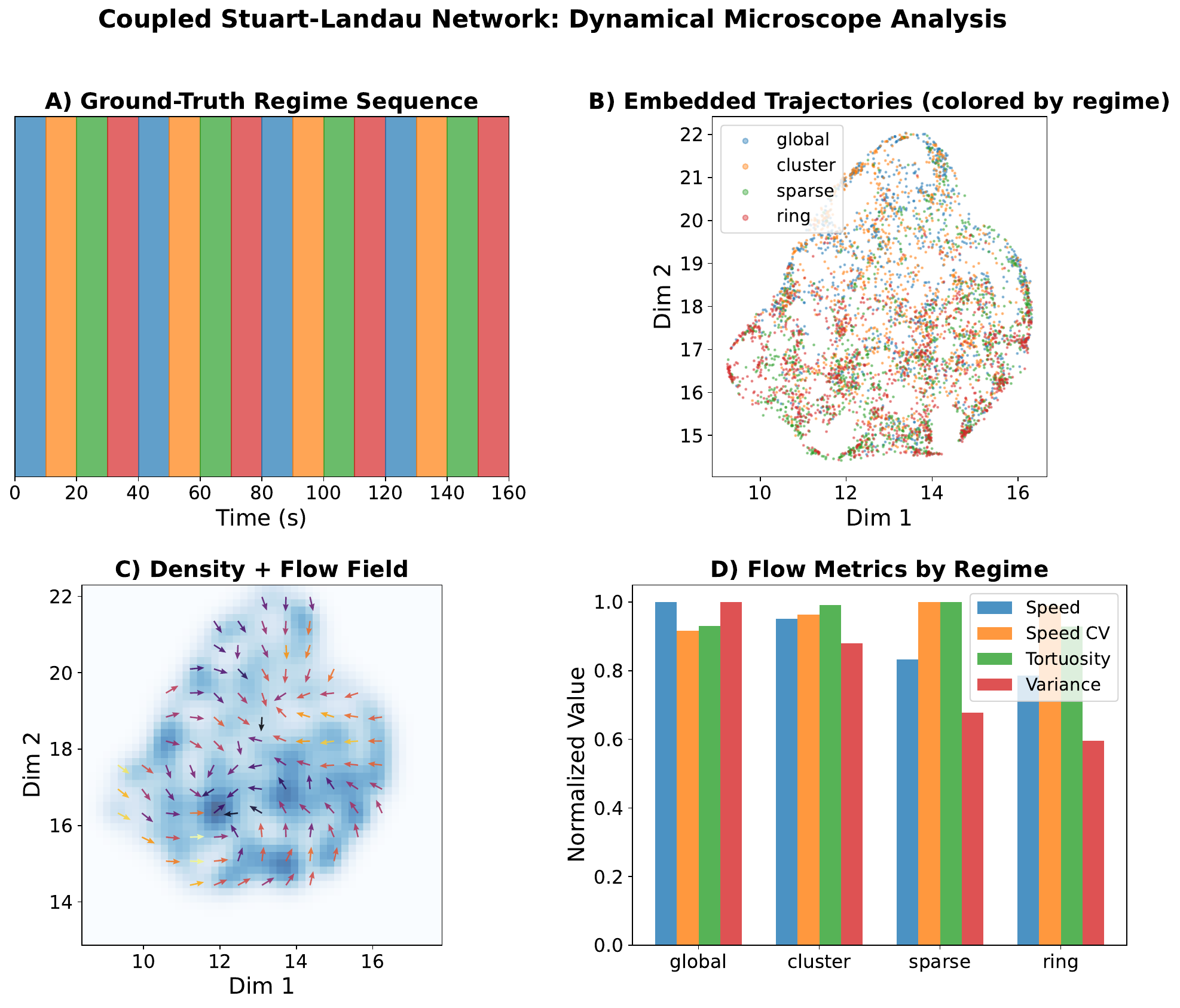}
    \caption{\textbf{Dynamical microscope analysis of coupled Stuart--Landau oscillator network.} (A) Ground-truth regime sequence: four coupling topologies (global, cluster, sparse, ring) cycling every 10~s for 160~s total. (B) UMAP-embedded latent trajectories colored by regime; regimes show substantial spatial overlap but visible clustering. (C) Density and flow field visualization showing local drift patterns across the shared manifold. (D) Normalized flow metrics by regime, demonstrating topology-dependent differences in speed, speed variability, tortuosity, and explored variance.}
    \label{fig:main_analysis}
\end{figure}

Figure~\ref{fig:main_analysis} summarizes the main validation results. Panel~A shows the ground-truth regime sequence, with rapid 10-second switching between topologies. Panel~B displays the UMAP-embedded latent trajectories colored by regime, demonstrating that while regimes occupy substantially overlapping regions of the embedding space, regime-specific clustering is still apparent.

Key observations:
\begin{itemize}
    \item \textbf{Spatial overlap}: Latent trajectories from different regimes occupy substantially overlapping regions, confirming that regime identity cannot be recovered from position alone.
    \item \textbf{Flow field structure}: Despite spatial overlap, the combined flow field (Panel~C) reveals coherent directional patterns, with local drift directions and magnitudes varying across the manifold.
    \item \textbf{Metric discrimination}: Panel~D shows that dynamical metrics successfully distinguish regimes, with global coupling producing highest speed and explored variance, while ring topology produces lowest values.
\end{itemize}

\subsection{Quantitative regime discriminability}

Table~\ref{tab:discriminability} summarizes the discriminability analysis across metrics. Speed and explored variance show large effect sizes ($\eta^2 > 0.5$), indicating that these metrics strongly differentiate the four coupling topologies. Tortuosity shows negligible discriminability ($\eta^2 < 0.01$), suggesting it captures aspects of trajectory geometry orthogonal to the topology contrasts in this simulation.

\begin{table}[h]
\centering
\caption{\textbf{Regime discriminability for coupled oscillator simulation.} One-way ANOVA results across four topology regimes (global, cluster, sparse, ring). $\eta^2 > 0.14$ indicates large effect.}
\label{tab:discriminability}
\small
\begin{tabular}{lccc}
\toprule
\textbf{Metric} & \textbf{$F$-statistic} & \textbf{$p$-value} & \textbf{$\eta^2$} \\
\midrule
Mean Speed & 67.8 & $4.2 \times 10^{-30}$ & 0.509 (large) \\
Explored Variance & 70.3 & $6.7 \times 10^{-31}$ & 0.518 (large) \\
Tortuosity & 0.61 & 0.61 & 0.009 (small) \\
\bottomrule
\end{tabular}
\end{table}

\subsection{Regime-specific flow metrics}

Table~\ref{tab:regime_metrics} presents detailed flow metrics for each coupling topology. The global topology produces the fastest dynamics (speed = 4.14) and broadest state-space exploration (variance = 26.7), consistent with all-to-all coupling promoting rapid collective motion. In contrast, the ring topology produces the slowest dynamics (speed = 3.26) and most confined trajectories (variance = 15.9), reflecting the local nature of ring coupling that restricts information propagation.

\begin{table}[h]
\centering
\caption{\textbf{Flow metrics by coupling topology.} Mean values computed over all windows within each regime.}
\label{tab:regime_metrics}
\small
\begin{tabular}{lcccc}
\toprule
\textbf{Metric} & \textbf{Global} & \textbf{Cluster} & \textbf{Sparse} & \textbf{Ring} \\
\midrule
Mean Speed & 4.14 & 3.94 & 3.45 & 3.26 \\
Speed CV & 0.27 & 0.29 & 0.30 & 0.29 \\
Tortuosity & 72.0 & 76.7 & 77.3 & 71.8 \\
Expl. Variance & 26.7 & 23.5 & 18.1 & 15.9 \\
Occ. Entropy & 5.33 & 5.21 & 5.11 & 4.98 \\
\bottomrule
\end{tabular}
\end{table}

The cluster and sparse topologies produce intermediate values, with sparse connectivity showing reduced speed and exploration compared to the highly connected global and cluster configurations.

\subsection{Regime-specific flow fields}

\begin{figure}[h]
    \centering
    \includegraphics[width=0.48\textwidth]{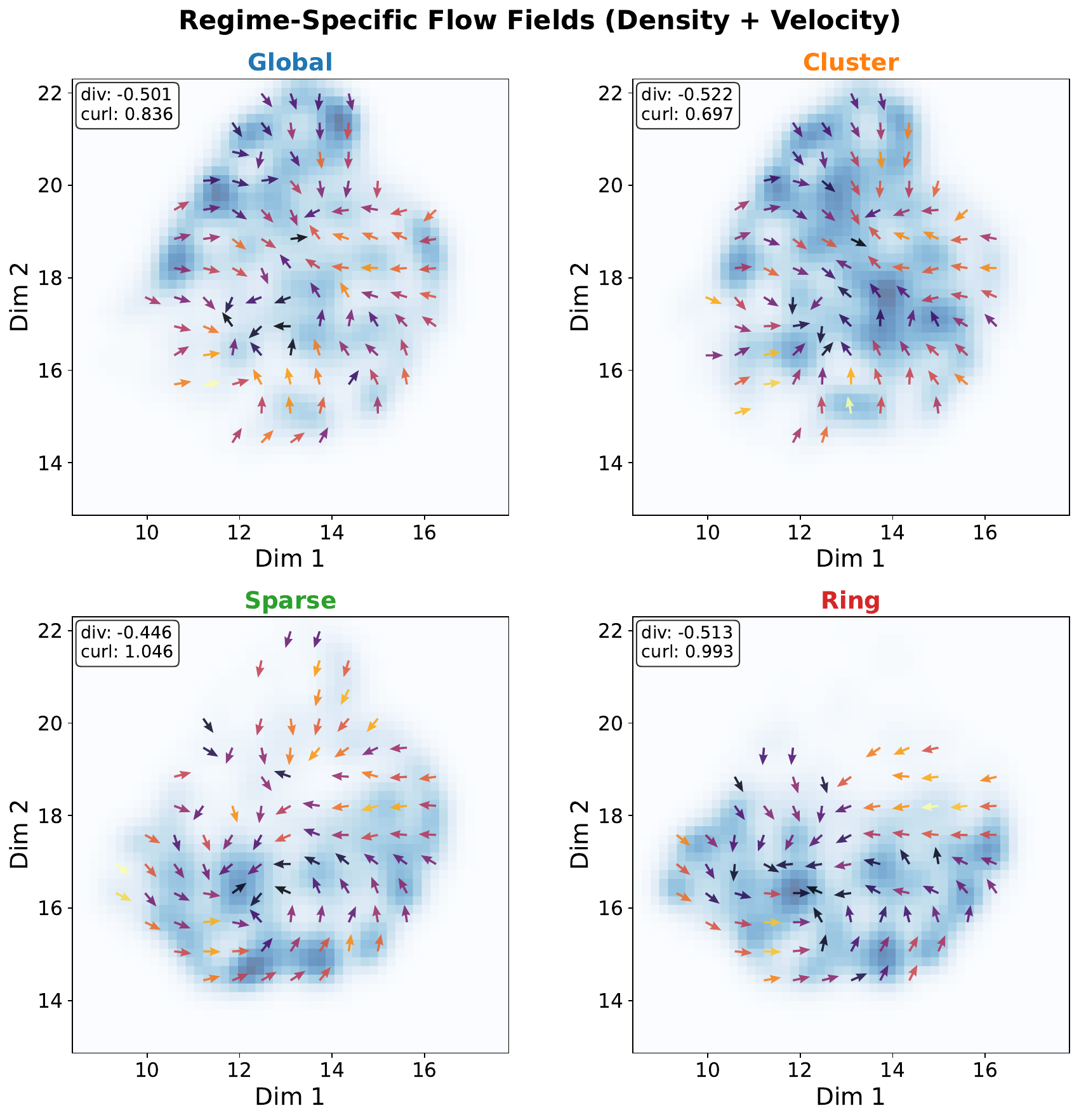}
    \caption{\textbf{Regime-specific flow fields.} Each panel shows the density (background) and flow field (arrows) for trajectories within a single coupling topology. Despite substantial spatial overlap, regimes exhibit distinct local flow patterns. Divergence and curl values are annotated for each regime.}
    \label{fig:flow_fields}
\end{figure}

Figure~\ref{fig:flow_fields} shows regime-specific flow field visualizations, revealing how local drift patterns differ across coupling topologies despite occupying overlapping regions of the embedding space.

Field-level metrics reveal additional structure:
\begin{itemize}
    \item \textbf{Divergence}: All regimes show negative mean divergence ($-0.44$ to $-0.52$), indicating overall contractive flow toward high-occupancy regions.
    \item \textbf{Curl/circulation}: The ring topology shows the highest curl magnitude and total circulation (16.5), consistent with its directional coupling structure promoting rotational flow patterns. Global coupling shows lowest circulation (5.9).
\end{itemize}

\subsection{Discriminability analysis}

\begin{figure}[h]
    \centering
    \includegraphics[width=0.48\textwidth]{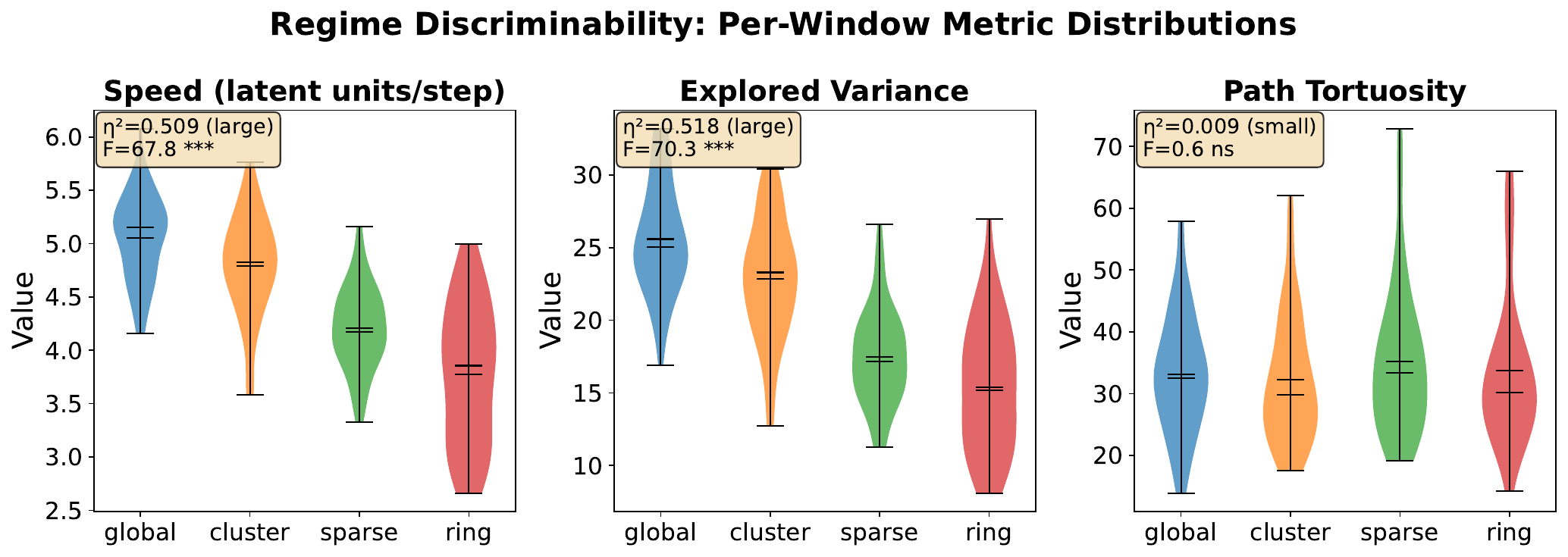}
    \caption{\textbf{Regime discriminability: per-window metric distributions.} Violin plots show the distribution of speed, variance, and tortuosity across non-overlapping windows for each regime. Effect sizes ($\eta^2$) and significance levels are annotated. Speed and variance show clear regime separation with large effect sizes, while tortuosity distributions overlap substantially.}
    \label{fig:discriminability}
\end{figure}

Figure~\ref{fig:discriminability} presents violin plots of per-window metric distributions, providing a visual summary of regime separability beyond mean differences.

The violin plots confirm that:
\begin{itemize}
    \item Speed distributions are well-separated, with global showing highest values and ring lowest.
    \item Explored variance shows similar separation, with a clear ordering from global (highest) through cluster and sparse to ring (lowest).
    \item Tortuosity distributions overlap almost completely, explaining its low discriminability for this topology contrast.
\end{itemize}

\subsection{Latent kinetic energy as a dynamical activity proxy}

While speed captures the average rate of motion through state space, we can derive a complementary quantity that emphasizes intermittency and burst-like dynamics: the latent kinetic energy proxy, defined as
\begin{equation}
E_k(t) = \|\mathbf{v}(t)\|^2,
\end{equation}
where $\mathbf{v}(t)$ is the velocity vector in latent space. Quadratic velocity functionals like $\|v\|^2$ have been used in dynamical systems to characterize activity intensity and intermittency independently of position, particularly in turbulence and stochastic flows \citep{Frisch1995,Falkovich2001}, making them a natural complement to first-order speed measures. Although mathematically related to speed, kinetic energy emphasizes rare high-velocity events that are averaged out in mean speed, and its distributional properties (skewness, kurtosis, tail structure) capture dynamical intermittency not reflected in first-order statistics alone.

\paragraph{Interpretation.}
This quantity is explicitly \emph{not} a measure of metabolic energy consumption---scalp signals do not provide direct access to cellular energetics. Rather, $E_k(t)$ serves as a dynamical activity proxy that can be interpreted as ``reconfiguration effort'' or ``state-space mobility'': how vigorously the system traverses its manifold at each moment.

\begin{figure}[h]
    \centering
    \includegraphics[width=0.48\textwidth]{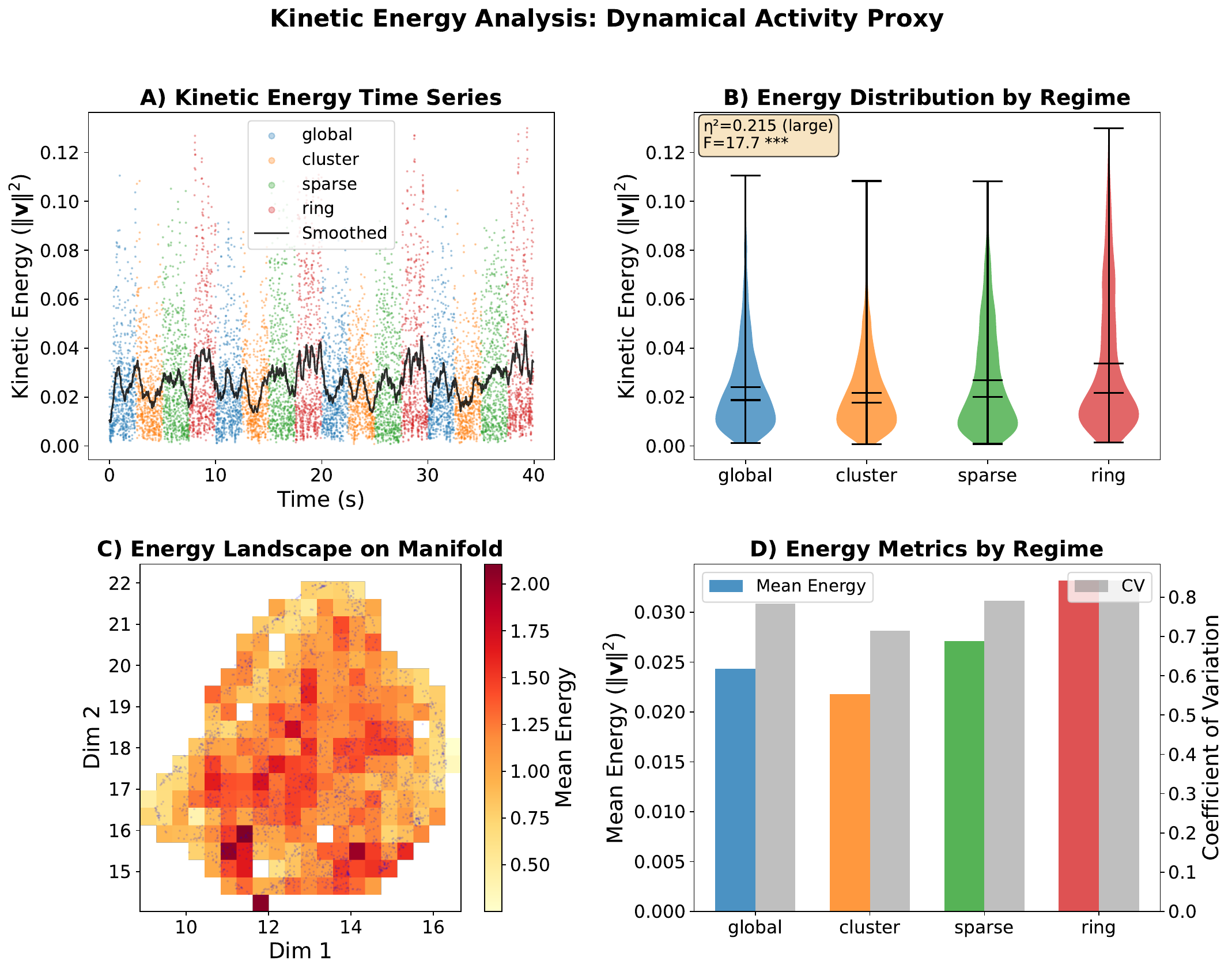}
    \caption{\textbf{Latent kinetic energy as a dynamical activity proxy.} (A) Kinetic energy time series $E_k(t) = \|\mathbf{v}(t)\|^2$ colored by regime, with smoothed trend (black). (B) Per-regime energy distributions showing systematic shifts and heavy tails. (C) Energy landscape: spatial distribution of mean $E_k$ on the 2D embedding, revealing regions of high vs.\ low dynamical activity. (D) Summary metrics by regime: mean energy and coefficient of variation (CV).}
    \label{fig:kinetic_energy}
\end{figure}

Figure~\ref{fig:kinetic_energy} presents the kinetic energy analysis. Panel~A shows the energy time series colored by regime, revealing that regime structure is visible as changes in variability and burstiness, not just mean level---consistent with intermittent dynamics featuring episodes of high activity embedded in quieter motion. Panel~B displays the per-regime energy distributions, showing systematic shifts with heavy tails indicating intermittent bursts. Panel~C presents the energy landscape: $E_k$ is computed in the full 32-dimensional latent space, then mean values are mapped to bins of the 2D visualization embedding to reveal spatial structure. This shows that some regions of the manifold are energetically more demanding to traverse than others. Panel~D summarizes mean energy and coefficient of variation by regime.

Table~\ref{tab:kinetic_energy} reports the kinetic energy metrics by regime, computed from the same velocity vectors as mean speed (ensuring $\mathbb{E}[E_k] = \mathbb{E}[\|v\|^2]$). The ring topology shows the highest mean energy ($E_k = 0.033$) and substantial variability (CV = 0.84), indicating rapid, intermittent state-space traversal. The cluster topology shows the lowest energy ($E_k = 0.022$) and lowest variability (CV = 0.72), consistent with stable, ``resting'' dynamics. Discriminability analysis yields $\eta^2 = 0.21$ (large effect), confirming that kinetic energy captures regime-dependent structure.

\begin{table}[h]
\centering
\caption{\textbf{Kinetic energy metrics by coupling topology.} Mean values computed from the same velocity source as speed metrics. Q95/Q50 ratio indicates tail heaviness (higher = more intermittent bursts).}
\label{tab:kinetic_energy}
\small
\begin{tabular}{lcccc}
\toprule
\textbf{Metric} & \textbf{Global} & \textbf{Cluster} & \textbf{Sparse} & \textbf{Ring} \\
\midrule
Mean speed $\|v\|$ & 0.145 & 0.139 & 0.152 & 0.167 \\
Mean $E_k = \|v\|^2$ & 0.024 & 0.022 & 0.027 & 0.033 \\
CV($E_k$) & 0.78 & 0.72 & 0.79 & 0.84 \\
Q95/Q50($E_k$) & 3.47 & 2.94 & 3.47 & 4.29 \\
\bottomrule
\end{tabular}
\vspace{0.5em}

\footnotesize{Discriminability for $E_k$: $F = 17.7$, $p = 3.3 \times 10^{-10}$, $\eta^2 = 0.21$ (large).}
\end{table}

The regime ordering is consistent between speed and kinetic energy (ring $>$ sparse $>$ global $>$ cluster), as expected when both are derived from the same velocity vectors. However, the Q95/Q50 ratio reveals that ring topology exhibits the heaviest tails (4.29), indicating more frequent high-energy bursts compared to cluster topology (2.94). This tail structure---captured by kinetic energy but obscured in mean speed---reflects the intermittent dynamics characteristic of ring coupling, where traveling wave patterns produce episodic acceleration events.

\paragraph{Note on Tables~\ref{tab:regime_metrics} vs.\ \ref{tab:kinetic_energy}.}
The speed values in Table~\ref{tab:regime_metrics} (flow metrics) were computed after standardization of the latent space, while Table~\ref{tab:kinetic_energy} reports raw latent velocities. Both orderings are consistent within their respective analyses; the absolute magnitudes differ due to the normalization.

\section{Discussion}
\label{sec:discussion}

\subsection{Summary of main findings}

This work introduced a ``dynamical microscope'' for multivariate oscillatory signals: a pipeline that transforms multichannel signals into circular phase--amplitude features, learns a compact latent representation with an autoencoder, and quantifies dynamical regimes using trajectory geometry and flow field statistics.

Validation on a coupled Stuart--Landau oscillator network with topology-switching demonstrated the central methodological point: regime differences expressed through coupling topology---rather than through disjoint state-space regions---are recoverable through trajectory-based analysis. The framework successfully recovered:
\begin{itemize}
    \item Topology-dependent differences in latent trajectory speed ($\eta^2 = 0.51$)
    \item Systematic variation in explored variance across coupling structures ($\eta^2 = 0.52$)
    \item Regime-specific flow field patterns including differential curl/circulation
    \item Complementary kinetic energy structure capturing dynamical intermittency ($\eta^2 = 0.21$)
\end{itemize}

Different metrics showed different sensitivity profiles: speed and variance strongly discriminated topologies, kinetic energy captured complementary intermittency structure, while tortuosity showed negligible discriminability for this particular contrast. This suggests that metric selection should be guided by the hypothesized nature of regime differences.

\subsection{From state occupancy to trajectory laws: what the microscope adds}

Analysis pipelines for complex signals typically either average over time (power spectra, connectivity) or discretize dynamics into state labels (microstates, HMM) and compare dwell times. These approaches privilege \emph{what} states occur over \emph{how} trajectories evolve between them. Our results motivate a complementary emphasis: regime differences appear as systematic changes in motion (speed, traversal geometry, flow field structure) rather than positional separation---consistent with a metastability view in which conditions preserve accessible configurations yet differ in transition dynamics.

A conceptual advantage of the dynamical microscope is the joint use of Lagrangian and Eulerian summaries. Time-ordered trajectories capture transient realizations, while density and flow field estimates characterize ensemble structure by pooling across time and/or realizations. This dual view is particularly appropriate for non-stationary signals, where stochasticity makes any single realization incomplete, and where condition differences may be expressed as shifts in the probability flow on a manifold rather than as separable clusters.

\subsection{Interpreting the topology contrast results}

The strong discriminability of speed and explored variance, combined with weak discriminability of tortuosity, has a natural interpretation in terms of the underlying oscillator dynamics:

\paragraph{Global coupling} promotes rapid collective motion because all oscillators directly influence each other, allowing perturbations to propagate instantaneously across the network. This produces high-speed, high-variance trajectories.

\paragraph{Ring coupling} restricts information flow to nearest neighbors, slowing collective dynamics and confining trajectories to smaller regions of state space. The directional nature of ring coupling also produces higher curl/circulation in the flow field.

\paragraph{Tortuosity} measures path curvature regardless of speed or extent. Since all four topologies produce comparably smooth trajectories (dominated by the oscillatory dynamics rather than coupling structure), tortuosity does not discriminate them effectively. This illustrates that different metrics capture different aspects of dynamical organization.

\subsection{Methodological considerations}

Several design choices improve interpretability:
\begin{enumerate}
    \item \textbf{Window-based statistical testing} using non-overlapping segments reduces short-timescale autocorrelation and supports descriptive ANOVA-based discriminability analysis.
    \item \textbf{Pooled embeddings} maintain a shared coordinate system for regime comparisons.
    \item \textbf{Primary metrics computed in full latent space} ensure projection invariance.
    \item \textbf{Field-level metrics} (divergence, curl) provide complementary characterization beyond point statistics.
\end{enumerate}

\subsection{Limitations}
\label{sec:limitations}

This study has several limitations that bound interpretation and motivate future work.

\paragraph{Synthetic validation only.}
The coupled oscillator network, while providing ground-truth regime structure, is a simplified model. Real-world systems may exhibit more complex dynamics, non-stationary noise, and regime structure not captured by topology-switching alone. However, the goal of this work is not to model neural data directly, but to validate a class of dynamical descriptors under conditions where ground truth is known; empirical applications are a necessary next step rather than a prerequisite for methodological soundness. Empirical validation on neural, physiological, or physical data is essential.

\paragraph{Single simulation configuration.}
Results are reported for one parameter configuration (coupling strength 5.0, 10-second regimes). Systematic exploration of parameter space would strengthen claims about robustness.

\paragraph{UMAP embedding choices.}
UMAP parameters affect visualization and flow field estimation. While primary metrics are computed in the full latent space, flow field visualizations depend on embedding quality.

\paragraph{Metric selection.}
Different metrics show different sensitivity profiles. The weak discriminability of tortuosity for topology contrasts does not mean tortuosity is uninformative in general---it may be highly sensitive to other types of regime differences (e.g., stability contrasts). Not all metrics discriminate all regime contrasts; the failure of tortuosity in this setting illustrates that different metrics probe orthogonal aspects of dynamics rather than providing redundant evidence.

\subsection{Future directions}

Key priorities for extending this work include:
\begin{enumerate}
    \item \textbf{Empirical validation}: Apply the framework to real-world multivariate signals where ground truth is unavailable but domain-specific hypotheses can be tested.
    \item \textbf{Parameter sensitivity}: Systematically explore how results depend on coupling strength, regime duration, and noise levels.
    \item \textbf{Alternative regime structures}: Test recovery of stability contrasts, bifurcation transitions, and other dynamical changes beyond topology-switching.
    \item \textbf{Multi-scale analysis}: Test whether flow signatures persist across temporal resolutions and embedding dimensions.
\end{enumerate}

\section{Conclusion}
\label{sec:conclusion}

We presented a dynamical microscope framework for multivariate oscillatory signals that emphasizes continuous trajectories, probability flow, and geometric organization in learned latent spaces. By combining circular phase--amplitude encoding with autoencoder-based representation learning and flow/geometry metrics, the approach characterizes dynamical regimes in terms of \emph{how} activity moves through state space, rather than relying primarily on discrete state labels or stationary averages.

Validation on a coupled Stuart--Landau oscillator network with topology-switching demonstrated that the framework recovers regime differences expressed through coupling structure, even when regimes occupy overlapping regions of state space. Speed and explored variance showed strong discriminability ($\eta^2 > 0.5$), while tortuosity captured aspects of trajectory geometry orthogonal to the topology contrast.

The dynamical microscope provides a general-purpose lens for studying non-stationary dynamics in multivariate time series. Future work will focus on empirical validation in neural, physiological, and physical systems, systematic parameter exploration, and extension to other types of dynamical regime structure.

\section*{Code and Data Availability}

The complete implementation of the dynamical microscope framework, including simulation code, analysis pipeline, and figure reproduction scripts, is available as an open-source Python package:

\begin{itemize}
    \item \textbf{Repository}: \url{https://github.com/furmanlukasz/flowprint}
    \item \textbf{Project page}: \url{https://furmanlukasz.github.io/flowprint}
    \item \textbf{Interactive environment}: GitHub Codespaces with pre-configured dependencies
\end{itemize}

\noindent The repository includes coupled Stuart--Landau oscillator simulation with configurable topologies, flow field estimation and trajectory metric computation, discriminability analysis, and scripts to reproduce all figures from this paper.

\section*{Acknowledgments}

\appendix

\begin{table}[h]
\centering
\caption{\textbf{Implementation summary.} Key pipeline parameters for reproducibility.}
\label{tab:implementation_summary}
\begin{tabular}{ll}
\toprule
\textbf{Parameter} & \textbf{Value} \\
\midrule
\multicolumn{2}{l}{\emph{Simulation}} \\
Sampling rate & 250~Hz \\
Duration & 160~s \\
Channels $C$ & 30 \\
Oscillators & 30 \\
\midrule
\multicolumn{2}{l}{\emph{Preprocessing}} \\
Bandpass filter & 1--30~Hz, 4th-order Butterworth, zero-phase \\
\midrule
\multicolumn{2}{l}{\emph{Hilbert extraction}} \\
Applied to & Continuous data per channel \\
\midrule
\multicolumn{2}{l}{\emph{Chunking}} \\
Chunk duration & 5.0~s (1250 samples) \\
\midrule
\multicolumn{2}{l}{\emph{Feature dimensions}} \\
Input features & $3C$ (cos$\phi$, sin$\phi$, log-amp) \\
\midrule
\multicolumn{2}{l}{\emph{Autoencoder}} \\
Latent dim $H$ & 32 \\
Training epochs & 50 \\
\midrule
\multicolumn{2}{l}{\emph{Velocity \& flow}} \\
Velocity estimation & Savitzky--Golay derivative \\
SG parameters & window=5, polyorder=2 \\
Flow field bins & $15 \times 15$ grid \\
Min occupancy & $k \geq 3$ points/bin \\
\midrule
\multicolumn{2}{l}{\emph{Embedding}} \\
Visualization space & UMAP \\
Dimensions & 2 (for visualization) \\
\midrule
\multicolumn{2}{l}{\emph{Statistical analysis}} \\
Window size & 50 samples \\
Test & One-way ANOVA \\
Effect size & $\eta^2$ \\
\bottomrule
\end{tabular}
\end{table}

\section{Robustness Analyses}
\label{app:robustness}

\subsection{Velocity estimation}
\label{app:velocity_robustness}

Velocity estimation is inherently scale-dependent: finite-difference estimates at short time scales ($\Delta t = 1\text{--}2$ samples) are dominated by high-frequency fluctuations, while coarser scales or smoothed derivatives reflect mesoscale flow. We tested finite differences ($\Delta t \in \{1, 2, 3, 5\}$) and Savitzky--Golay derivatives (window $\in \{5, 7, 9\}$).

At all configurations, effect directions were consistent across regime comparisons. Effect magnitudes varied with temporal scale, as expected. SG($w = 5$) was selected for primary analyses as it provides noise suppression while preserving trajectory curvature.

\bibliographystyle{elsarticle-harv}

\begin{thebibliography}{00}

\bibitem[Kelso(1995)]{Kelso1995}
J.~A.~S. Kelso.
\newblock \emph{Dynamic Patterns: The Self-Organization of Brain and Behavior}.
\newblock MIT Press, Cambridge, MA, 1995.

\bibitem[Rabinovich et~al.(2008)]{Rabinovich2008}
M.~I. Rabinovich, R.~Huerta, P.~Varona, and V.~S. Afraimovich.
\newblock Transient cognitive dynamics, metastability, and decision making.
\newblock \emph{PLoS Computational Biology}, 4(5):e1000072, 2008.

\bibitem[Tognoli and Kelso(2014)]{TognoliKelso2014}
E.~Tognoli and J.~A.~S. Kelso.
\newblock The metastable brain.
\newblock \emph{Neuron}, 81(1):35--48, 2014.

\bibitem[Breakspear(2017)]{Breakspear2017}
M.~Breakspear.
\newblock Dynamic models of large-scale brain activity.
\newblock \emph{Nature Neuroscience}, 20(3):340--352, 2017.

\bibitem[Deco et~al.(2017)]{Deco2017}
G.~Deco, M.~L. Kringelbach, V.~K. Jirsa, and P.~Ritter.
\newblock The dynamics of resting fluctuations in the brain: metastability and its dynamical cortical core.
\newblock \emph{Scientific Reports}, 7:3095, 2017.

\bibitem[Michel and Koenig(2018)]{Michel2018}
C.~M. Michel and T.~Koenig.
\newblock EEG microstates as a tool for studying the temporal dynamics of whole-brain neuronal networks: a review.
\newblock \emph{NeuroImage}, 180:577--593, 2018.

\bibitem[Baker et~al.(2014)]{Baker2014}
A.~P. Baker, M.~J. Brookes, I.~Rezek, S.~M. Smith, T.~Behrens, J.~Probert, and M.~W. Woolrich.
\newblock Fast transient networks in spontaneous human brain activity.
\newblock \emph{eLife}, 3:e01867, 2014.

\bibitem[Vidaurre et~al.(2018)]{Vidaurre2018}
D.~Vidaurre, L.~T. Hunt, A.~J. Quinn, B.~A.~E. Hunt, M.~J. Brookes, A.~C. Nobre, and M.~W. Woolrich.
\newblock Discovering dynamic brain networks from big data in rest and task.
\newblock \emph{NeuroImage}, 180:646--656, 2018.

\bibitem[Banville et~al.(2021)]{Banville2021}
H.~Banville, O.~Chehab, A.~Hyv{\"a}rinen, D.~A. Engemann, and A.~Gramfort.
\newblock Uncovering the structure of clinical EEG signals with self-supervised learning.
\newblock \emph{Journal of Neural Engineering}, 18(4):046020, 2021.

\bibitem[Kostas et~al.(2021)]{Kostas2021}
D.~Kostas, S.~Aroca-Ouellette, and F.~Rudzicz.
\newblock BENDR: Using transformers and a contrastive self-supervised learning task to learn from massive amounts of EEG data.
\newblock \emph{Frontiers in Human Neuroscience}, 15:653659, 2021.

\bibitem[Marwan et~al.(2007)]{Marwan2007}
N.~Marwan, M.~C. Romano, M.~Thiel, and J.~Kurths.
\newblock Recurrence plots for the analysis of complex systems.
\newblock \emph{Physics Reports}, 438(5--6):237--329, 2007.

\bibitem[Cavanna et~al.(2018)]{Cavanna2018}
F.~Cavanna, M.~G. Vilas, M.~Palmucci, and E.~Tagliazucchi.
\newblock Dynamic functional connectivity and brain metastability during altered states of consciousness.
\newblock \emph{NeuroImage}, 180:383--395, 2018.

\bibitem[Boashash(1992)]{Boashash1992}
B.~Boashash.
\newblock Estimating and interpreting the instantaneous frequency of a signal.
\newblock \emph{Proceedings of the IEEE}, 80(4):520--538, 1992.

\bibitem[Canolty et~al.(2006)]{Canolty2006}
R.~T. Canolty, E.~Edwards, S.~S. Dalal, M.~Soltani, D.~Nagarajan, H.~E. Kirsch, M.~S. Berger, N.~M. Barbaro, and R.~T. Knight.
\newblock High gamma power is phase-locked to theta oscillations in human neocortex.
\newblock \emph{Science}, 313(5793):1626--1628, 2006.

\bibitem[Tort et~al.(2010)]{Tort2010}
A.~B.~L. Tort, R.~Komorowski, H.~Eichenbaum, and N.~Kopell.
\newblock Measuring phase-amplitude coupling between neuronal oscillations of different frequencies.
\newblock \emph{Journal of Neurophysiology}, 104(2):1195--1210, 2010.

\bibitem[Hinton and Salakhutdinov(2006)]{Hinton2006}
G.~E. Hinton and R.~R. Salakhutdinov.
\newblock Reducing the dimensionality of data with neural networks.
\newblock \emph{Science}, 313(5786):504--507, 2006.

\bibitem[Cunningham and Yu(2014)]{CunninghamYu2014}
J.~P. Cunningham and B.~M. Yu.
\newblock Dimensionality reduction for large-scale neural recordings.
\newblock \emph{Nature Neuroscience}, 17(11):1500--1509, 2014.

\bibitem[Brunton et~al.(2016)]{Brunton2016}
S.~L. Brunton, J.~L. Proctor, and J.~N. Kutz.
\newblock Discovering governing equations from data by sparse identification of nonlinear dynamical systems.
\newblock \emph{Proceedings of the National Academy of Sciences}, 113(15):3932--3937, 2016.

\bibitem[Chen et~al.(2018)]{Chen2018}
R.~T.~Q. Chen, Y.~Rubanova, J.~Bettencourt, and D.~Duvenaud.
\newblock Neural ordinary differential equations.
\newblock In \emph{Advances in Neural Information Processing Systems (NeurIPS)}, 2018.

\bibitem[Kingma and Welling(2014)]{KingmaWelling2014}
D.~P. Kingma and M.~Welling.
\newblock Auto-encoding variational Bayes.
\newblock In \emph{International Conference on Learning Representations (ICLR)}, 2014.

\bibitem[Pandarinath et~al.(2018)]{Pandarinath2018}
C.~Pandarinath, D.~J. O'Shea, J.~Collins, et al.
\newblock Inferring single-trial neural population dynamics using sequential auto-encoders.
\newblock \emph{Nature Methods}, 15(10):805--815, 2018.

\bibitem[Savitzky and Golay(1964)]{SavitzkyGolay1964}
A.~Savitzky and M.~J.~E. Golay.
\newblock Smoothing and differentiation of data by simplified least squares procedures.
\newblock \emph{Analytical Chemistry}, 36(8):1627--1639, 1964.

\bibitem[Schmid(2010)]{Schmid2010DMD}
P.~J. Schmid.
\newblock Dynamic mode decomposition of numerical and experimental data.
\newblock \emph{Journal of Fluid Mechanics}, 656:5--28, 2010.

\bibitem[Linderman et~al.(2017)]{Linderman2017rSLDS}
S.~W. Linderman, M.~J. Johnson, A.~C. Miller, R.~P. Adams, D.~M. Blei, L.~Paninski.
\newblock Bayesian learning and inference in recurrent switching linear dynamical systems.
\newblock In \emph{Proceedings of the 20th International Conference on Artificial Intelligence and Statistics (AISTATS)}, PMLR 54:914--922, 2017.

\bibitem[Cohen(1988)]{Cohen1988}
J.~Cohen.
\newblock \emph{Statistical Power Analysis for the Behavioral Sciences}.
\newblock 2nd ed., Lawrence Erlbaum Associates, Hillsdale, NJ, 1988.

\bibitem[Frisch(1995)]{Frisch1995}
U.~Frisch.
\newblock \emph{Turbulence: The Legacy of A.~N.~Kolmogorov}.
\newblock Cambridge University Press, 1995.

\bibitem[Falkovich et~al.(2001)]{Falkovich2001}
G.~Falkovich, K.~Gaw\c{e}dzki, and M.~Vergassola.
\newblock Particles and fields in fluid turbulence.
\newblock \emph{Reviews of Modern Physics}, 73(4):913--975, 2001.

\bibitem[Strogatz(1994)]{Strogatz1994}
S.~H. Strogatz.
\newblock \emph{Nonlinear Dynamics and Chaos}.
\newblock Perseus Books, 1994.

\bibitem[McInnes et~al.(2018)]{McInnes2018}
L.~McInnes, J.~Healy, and J.~Melville.
\newblock UMAP: Uniform Manifold Approximation and Projection for Dimension Reduction.
\newblock \emph{arXiv:1802.03426}, 2018.

\end{thebibliography}

\end{document}